\documentclass[ALICE,manyauthors]{cernphprep}

\usepackage{datetime}
\usepackage{amsmath}
\usepackage{xcolor}
\usepackage[comma,square,numbers,sort&compress]{natbib}
\usepackage{hyperref}
\usepackage{lineno}
\usepackage{multirow}
\usepackage{graphicx}
\usepackage{placeins}
\usepackage[T1]{fontenc}
\usepackage{orcidlink}

\begin{document}%

\begin{titlepage}
\PHyear{2022}
\PHnumber{210}       
\PHdate{11 October}  

%

\title{$\mathbf{\psi(2S)}$ suppression in Pb--Pb collisions at the LHC}
\ShortTitle{$\psi(2S)$ suppression in Pb--Pb collisions at the LHC}  

\Collaboration{ALICE Collaboration\thanks{See Appendix~\ref{app:collab} for the list of collaboration members}}
\ShortAuthor{ALICE Collaboration} 


\begin{abstract}
The production of the $\psi(2S)$ charmonium state was measured with ALICE in Pb--Pb collisions at $\sqrt{s_{\rm NN}}=5.02$ TeV, in the dimuon decay channel. A significant signal was observed for the first time at LHC energies down to zero transverse momentum, at forward rapidity ($2.5<y<4$). The measurement of the ratio of the inclusive production cross sections of the $\psi(2S)$ and J/$\psi$ resonances is reported as a function of the centrality of the collisions and of transverse momentum, in the region $p_{\rm T}<12$ GeV/$c$. The results are compared with the corresponding measurements in pp collisions, by forming the double ratio $[\sigma^{\psi(2S)}/\sigma^{J/\psi}]_{\rm{Pb-Pb}}/[\sigma^{\psi(2S)}/\sigma^{J/\psi}]_{\rm{pp}}$. It is found that in Pb--Pb collisions the $\psi(2S)$ is suppressed by a factor of $\sim 2$ with respect to the J/$\psi$.
The $\psi(2S)$ nuclear modification factor $R_{\rm AA}$ was also obtained as a function of both centrality and $p_{\rm T}$.
The results show that the $\psi(2S)$ resonance yield is strongly suppressed in Pb--Pb collisions, by a factor up to $\sim 3$ with respect to pp.  Comparisons of cross section ratios with previous SPS findings by the NA50 experiment and of $R_{\rm AA}$ with higher-$p_{\rm T}$ results at LHC energy are also reported.  These results and the corresponding comparisons with calculations of transport and statistical models address  questions on the presence and properties of charmonium states in the quark--gluon plasma formed in nuclear collisions at the LHC. 

\end{abstract}
\end{titlepage}
\setcounter{page}{2}
Quarkonia, the bound states of a heavy quark--antiquark pair, represent an important test bench for quantum chromodynamics (QCD), the theory of strong interactions~\cite{Brambilla:2010cs}. The production process of the pair is governed by the hard scale corresponding to the quark mass and occurs on a very short time ($\sim 0.1$ fm/$c$), while its binding is a soft process, characterized by timescales that can be an order of magnitude larger~\cite{Kharzeev:1999bh,Hufner:2000jb}. 
Static properties of quarkonia, and in particular their complex spectroscopy, can be reproduced by formulating QCD on a discrete lattice in space and time~\cite{Davies:2003ik}. The quarkonium states can also be used as a probe of strongly interacting extended systems, and in particular of the quark--gluon plasma (QGP), a state of matter where quarks and gluons are deconfined over distances much larger than the hadronic size ($\sim 1$~fm). In such a state, which can be created by collisions of heavy ions at ultrarelativistic energies, the large density of free color charges leads to a screening of the quark--antiquark binding and to the dissociation of quarkonia~\cite{Matsui:1986dk}. Effects related to a collisional damping of the states can also be present, leading to a loss of correlation in the pair and, consequently, to a modification of the spectral functions in the QGP~\cite{Laine:2006ns}. Finally, the QGP created in the collision expands and cools down until it crosses the pseudocritical temperature $T_{\rm c}$ (of about 157 MeV for a system with zero net baryonic number~\cite{HotQCD:2018pds,Borsanyi:2020fev}) for the transition to a hadronic phase. Close to this transition, if the initial multiplicity of heavy quark pairs is large, recombination processes, which counterbalance to a certain extent the suppression in the QGP, may become sizable~\cite{BraunMunzinger:2000px,Thews:2000rj}; i.e., quarks and antiquarks close in phase space can recombine to form a quarkonium state. These processes may already be effective even in the QGP phase~\cite{Grandchamp:2003uw}.

A further important element in the study of quarkonium production in heavy-ion collisions is their rich variety of states. Restricting the discussion to charmonia, bound states of charm--anticharm quarks, the ground-level J/$\psi$ vector meson and its radial excitation $\psi({\rm 2S})$ differ in binding energy by more than a factor of 10 ($\sim 640$ versus $\sim 50$ MeV, respectively) and by about a factor of 2 in size~\cite{Karsch:1990wi,Satz:2005hx}. As a consequence, 
the dissociation of the charmonium states depends on
the temperature of the medium and is expected to occur sequentially, reflecting the increasing values of their binding energies~\cite{Digal:2001ue}.
Also, the recombination processes might, in principle, have different features, with the larger-size charmonium states being produced later in the evolution of the system~\cite{Du:2015wha}.  Current theoretical approaches for the complex phenomenology of charmonium production in nuclear collisions include transport models~\cite{Zhao:2011cv,Zhou:2014kka}, where dissociation and recombination rates for quarkonium states in the QGP are calculated taking into account a lattice-QCD inspired evaluation of the dependence of their spectral properties on the evolving thermodynamic properties of the medium. 
In the statistical hadronization model (SHMc)~\cite{Andronic:2019wva}, charmonia are assumed to be formed at hadronization  according to statistical weights and introducing a charm fugacity factor related to charm conservation and determined by the charm production cross section\footnote{In the frame of SHMc it can be more appropriate to use the word ``combination'' rather than ``recombination''
as there is no binding of charmonium states in the QGP phase.}. The availability of accurate experimental results for various charmonium states represents a crucial input for the evaluation of the theory approaches and ultimately for the understanding of the existence of bound states of heavy quarks in the QGP. 

On the experimental side, a suppression of the J/$\psi$ in Pb--Pb collisions was observed by the NA50 experiment at the CERN Super Proton Synchrotron (SPS)~\cite{Alessandro:2004ap} (center of mass energy per nucleon--nucleon collision $\sqrt{s_{\rm NN}}=17.3$ GeV) and subsequently confirmed at the Relativistic Heavy Ion Collider (RHIC) by PHENIX~\cite{PHENIX:2011img} and STAR~\cite{STAR:2019fge} (Au--Au at $\sqrt{s_{\rm NN}}=200$ GeV). At the LHC (Pb--Pb at $\sqrt{s_{\rm NN}}=2.76$ and 5.02 TeV), the ALICE Collaboration has unambiguously demonstrated the existence and role of the recombination processes, by observing at low transverse momentum ($p_{\rm T}$), and in central collisions, a smaller suppression compared to lower-energy results~\cite{ALICE:2019nrq}. At high $p_{\rm T}$, CMS and ATLAS results indicate a strong J/$\psi$ suppression~\cite{Aaboud:2018quy,Sirunyan:2016znt}, reaching a value of $\sim 4$ for central collisions, where the geometric overlap of the colliding nuclei is maximal. The suppression in Pb--Pb collisions is quantified via the nuclear modification factor $R_{\rm AA}$, defined as the ratio between the J/$\psi$ yield in Pb--Pb and the product of the corresponding J/$\psi$ cross section in pp collisions times the average nuclear thickness function $\langle T_{\rm AA}\rangle$~\cite{ALICE-PUBLIC-2018-011}, a quantity proportional to the number of nucleon--nucleon collisions.  

The $\psi({\rm 2S})$ measurements are more challenging, due to the $\sim 7.5$ lower branching ratio to muon pairs with respect to the J/$\psi$, and the $\sim 6$ times smaller production cross section in pp collisions at LHC energy~\cite{Abelev:2014qha}. The most accurate result until today was obtained by NA50, which measured a decrease of the cross section ratio between $\psi({\rm 2S})$ and J/$\psi$ 
by a factor of $\sim 2$, in Pb--Pb collisions at $\sqrt{s_{\rm NN}}=17.3$ GeV~\cite{NA50:2006yzz}, when increasing the collision centrality. Both transport~\cite{Grandchamp:2003uw,Grandchamp:2004tn} and statistical hadronization models~\cite{Andronic:2006ky} were able to reproduce the main features of this result (see e.g. Fig. 37 of Ref.~\cite{Rapp:2009my}).
More recently, $\psi({\rm 2S})$ production was studied by ATLAS~\cite{Aaboud:2018quy} and CMS~\cite{Sirunyan:2016znt,Sirunyan:2017isk}, measuring the double ratio between the $\psi({\rm 2S})$ and J/$\psi$ cross sections in Pb--Pb and pp collisions at $\sqrt{s_{\rm NN}}=5.02$ TeV. Their analyses, carried out at  midrapidity,  
show in the  high-$p_{\rm T}$ region a strong relative suppression of the $\psi({\rm 2S})$ with respect to J/$\psi$, by a factor $\sim 2$. For a complete characterization of $\psi({\rm 2S})$ production at these energies, an extension of these results toward low $p_{\rm T}$, the kinematic region where recombination effects are maximal, is needed. To date, the previous result by ALICE~\cite{Adam:2015isa} at $\sqrt{s_{\rm NN}}=2.76$ TeV does not allow a firm conclusion, due to the large uncertainties.

In this Letter, we present results on inclusive $\psi({\rm 2S})$ production, obtained by ALICE in Pb--Pb collisions at $\sqrt{s_{\rm NN}}=5.02$ TeV. The $\psi({\rm 2S})$ is studied by means of its decay to muon pairs in the region $2.5<y<4$ and down to zero $p_{\rm T}$. Measurements of the (double) ratio of production cross sections between the $\psi({\rm 2S})$ and J/$\psi$, as well as of the $\psi({\rm 2S})$ $R_{\rm AA}$, are given. The ALICE detector is described extensively in Refs.~\cite{Aamodt:2008zz,Abelev:2014ffa}. In particular, muon detection is carried out by a forward spectrometer consisting of a 3~Tm dipole magnet, a system of two hadron absorbers, and five tracking (Cathode Pad Chambers) and two triggering (Resistive Plate Chambers) stations. 
The minimum-bias (MB) trigger is obtained as a coincidence of signals from the two V0 scintillator arrays ($-3.7<\eta < -1.7$ and $2.8 <\eta< 5.1$), also used for the rejection of beam--gas interactions and for the determination of the centrality of the collisions (see below).

The data analyzed in this Letter were collected in 2015 and 2018 and correspond to an integrated luminosity $L_{\rm int}\sim 750$ $\mu {\rm b}^{-1}$. The collisions were classified from central to peripheral according to the decreasing energy deposition in the V0 detectors, which can be related to the degree of geometric overlap of the colliding nuclei~\cite{ALICE:2013hur,ALICE-PUBLIC-2018-011}. 
Events were recorded using a dimuon trigger, given by the coincidence of a MB trigger together with the detection of a pair of particles with opposite charges  in the triggering system of the muon spectrometer. The trigger algorithm applies a non-sharp $p_{\rm T}$ threshold, which has 50\% efficiency at 1 GeV/$c$ and becomes fully efficient ($>98$\%) for $p_{\rm T}> 2$ GeV/$c$. Selection criteria were applied at the single muon and muon pair levels (see Ref.~\cite{ALICE:2018bdo} for details). 
Opposite-sign dimuons were selected in the rapidity interval  $2.5<y<4$.

The signal extraction procedure for the $\psi({\rm 2S})$ and  J/$\psi$ was based  on $\chi^2$ minimization fits of the opposite-sign dimuon invariant-mass spectra. The combinatorial background was subtracted with the help of an event mixing procedure, described in detail in Ref.~\cite{Adam:2015isa}. The resonance signals were described by a double-sided Crystal Ball function or a pseudo-Gaussian with a mass-dependent width~\cite{ALICE-Quarkonia-signal-extraction}. The position of the pole mass of the J/$\psi$, as well as its width, were kept as free parameters in the fitting procedure. For the $\psi({\rm 2S})$, due to the much smaller statistical significance of its signal, its mass was bound to that of the J/$\psi$, via the mass difference of the two resonances as provided by the Particle Data Group~\cite{Workman:2022ynf}, $m_{\psi({\rm 2S})}= m^{\rm FIT}_{\rm J/\psi} + (m^{\rm PDG}_{\psi({\rm 2S})} - m^{\rm PDG}_{\rm J/\psi})$. The width was also bound to that of the J/$\psi$, imposing the same ratio of the J/$\psi$ and $\psi({\rm 2S})$ widths as the one measured in the data sample of pp collisions at $\sqrt{s}=13$ TeV~\cite{ALICE:2017leg} or in Monte Carlo (MC) simulations (see details below). The same data sample, or the MC, was also used to fix the non-Gaussian tails of the resonance mass spectra. 
The continuum component of the correlated background remaining in the dimuon distributions after mixed-event subtraction and originating mainly from semi-muonic decays of pairs of heavy-flavor hadrons was parametrized using various empirical functions. 
Fits were performed in two invariant mass intervals, $2<m_{\mu\mu}<5$ GeV/$c^2$ and $2.2<m_{\mu\mu}<4.5$ GeV/$c^2$, roughly centered in the resonance region. The resonance signals were extracted in four $p_{\rm T}$ classes 
for the centrality range 0--90\% and in four centrality classes. 
For the latter series, the selections $p_{\rm T}<12$ GeV/$c$ and $0.3<p_{\rm T}<12$ GeV/$c$ were used for the two most central and peripheral classes, respectively, to remove the low-$p_{\rm T}$ contribution from photoproduction~\cite{ALICE:2022zso} which becomes important for peripheral collisions. 
For each kinematic and/or centrality selection, the numbers of detected $\psi({\rm 2S})$ and J/$\psi$ were obtained by averaging the results of the various fits. For the complete data sample (0--90\%) they amount to $1.3\times 10^4$ and $9.2\times 10^5$, respectively.
All the invariant mass spectra and the results of the corresponding fits are reported in Appendix~\ref{app:fits}.

The product of acceptance times efficiency ($A\times\epsilon$) has been calculated by means of a MC simulation. The $p_{\rm T}$ and $y$ distributions for the generated J/$\psi$ were matched to those extracted from data using an iterative procedure as done in Ref.~\cite{Acharya:2018kxc}, and the same distributions were also used for the $\psi({\rm 2S})$. The misalignment of the detection elements as well as the time-dependent status of each electronic channel during the data-taking period were taken into account in the simulation. The resonance signals were embedded into real events in order to properly
reproduce the effect of detector occupancy and its variation from one centrality class to another and then reconstructed. The centrality and $p_{\rm T}$-integrated $A\times\epsilon$ values, relative to the $2.5<y<4$ interval, are 13.5\% and 17.3\% for J/$\psi$ and $\psi({\rm 2S}$). As a function of $p_{\rm T}$ and centrality the $A\times\epsilon$ vary within a factor of $\sim 2$ and by about 5\%, respectively. 

The evaluation of the double ratios and the nuclear modification factors requires the measurement of the charmonium cross sections in pp collisions at $\sqrt{s}=5.02$ TeV. Results from Ref.~\cite{ALICE:2021qlw} were used for this purpose, by appropriately combining, where necessary, the $\psi({\rm 2S})$ cross sections and the ratios $\sigma_{\psi({\rm 2S})}/\sigma_{{\rm J}/\psi}$, in order to match the $p_{\rm T}$ binning used in this analysis. 

The normalization of the yields per event in Pb--Pb collisions is obtained calculating the number of ``equivalent'' minimum-bias events  as the product of the number of dimuon-triggered events ($\sim 4\times10^8$) times the inverse of the probability
of having a dimuon trigger in a MB event ($F$), following the procedure described e.g., in Ref.~\cite{ALICE:2020jff}. For the 0--90\% centrality  class, the value of the normalization factor is $F= 13.1 \pm 0.1$. 
Finally, the $\langle T_{\rm AA} \rangle$ values were taken from Ref.~\cite{ALICE-PUBLIC-2018-011} for the centrality intervals directly quoted there, or by combining their values for the other intervals.

\begin{table}[!ht]
	\begin{center}
		\caption{Contributions to the systematic uncertainties (in percentage). In each column, values with an asterisk correspond to the systematic uncertainties correlated as a function of the corresponding variable. The pp reference entry includes a 1.8\% luminosity uncertainty.}\label{tab:1}
		\begin{tabular}{c|c|c}
			\hline
			& versus centrality & versus $p_{\rm T}$\\
			\hline
			& \multicolumn{2}{c}{$\psi({\rm 2S})$ $R_{\rm AA}$} \\
			\hline
			Signal extraction & 16--22 & 12--25\\ 
			\hline
			Tracking effic. & 3 * & 3 \\
			\hline
			Trigger effic. & 1.6 * & 1.5--2\\
			\hline
			Matching effic. & 1 * & 1 \\
			\hline
			MC input & 2 * & 2 \\
			\hline
			$F$ & 0.7 * & 0.7 *\\
			\hline
			$\langle T_{\rm AA}\rangle$ & 0.7--2.3 & 1 *\\
			\hline
			Centrality & 0--7 & 0.3 *\\
			\hline
			pp ref. & 4.7 * & 7.9--11.1 \\
			\hline
			& \multicolumn{2}{c}{(Double) ratios} \\
			\hline
			Signal extraction & 16--23 & 12--24\\ 
			\hline
			pp ref. & 6.7 * & 5.5--8.8 \\
			\hline
		\end{tabular}
	\end{center}
\end{table}
	
A summary of the systematic uncertainties affecting the calculation of the (double) ratios and $\psi({\rm 2S})$ $R_{\rm AA}$ is given in Table~\ref{tab:1}. They were obtained following similar procedures as those adopted for previous charmonium analyses at forward rapidity, that are detailed e.g., in Refs.~\cite{Adam:2016rdg,Acharya:2019iur}. For the signal extraction, the systematic uncertainty was calculated as the root mean square of the values of the number of $\psi({\rm 2S})$ obtained by combining different fitting ranges, signal and residual background shapes. In addition, two different normalization ranges of the mixed-event spectra were tested. Fits with a $\chi^2/{\rm ndf}>2.5$ were excluded from the calculation. 
For the ingredients entering the $A\times\epsilon$ calculation, the systematic uncertainties related to tracking, triggering, and matching of candidate tracks between tracking and triggering detectors
were obtained by comparing information obtained in MC and in data, as described in Ref.~\cite{Acharya:2019iur}. The uncertainty due to the generated $y$ and $p_{\rm T}$ resonance shapes in the MC was estimated as in Ref.~\cite{Acharya:2018kxc}, taking into account the
statistical uncertainty on the measured distributions used for their definition in the iterative procedure and the possible correlations between the distributions in $y$ and $p_{\rm T}$. The uncertainty on $F$ was obtained by comparing the results of two different  evaluations as discussed in Ref.~\cite{ALICE:2020jff}, while for $\langle T_{\rm AA}\rangle$ they were computed by varying the input parameters of the Glauber calculation used for their estimate~\cite{ALICE-PUBLIC-2018-011}. For the centrality evaluation, the systematic uncertainty was obtained varying by $\pm1$\% the value of the V0 signal amplitude corresponding to the most central 90\% of the hadronic Pb–Pb cross section and re-extracting the resonance signal under this hypothesis, as detailed in Ref.~\cite{Adam:2016rdg}. For the pp reference, uncertainties were obtained by combining the corresponding values from the narrower $p_{\rm T}$ intervals reported in Ref.~\cite{ALICE:2021qlw}. A further uncertainty related to the evaluation of the pp luminosity is also considered in the $R_{\rm AA}$ evaluation~\cite{ALICE:2021qlw}. All the uncertainties discussed in this paragraph are added in quadrature to obtain the total systematic uncertainty.

When the $\psi({\rm 2S})$ to J/$\psi$ cross section ratios are computed, all uncertainties cancel out except the one related to the signal extraction which is dominated by the former resonance. For the double ratios, the uncertainties on the pp cross section ratio between $\psi({\rm 2S})$ and J/$\psi$ were also obtained starting from Ref.~\cite{ALICE:2021qlw}. All the results shown in this Letter and the corresponding model calculations refer to inclusive quarkonium production, which includes nonprompt quarkonia, originating from the decay of b hadrons.

\begin{figure}[h]
\begin{center}
\includegraphics[scale=0.50]{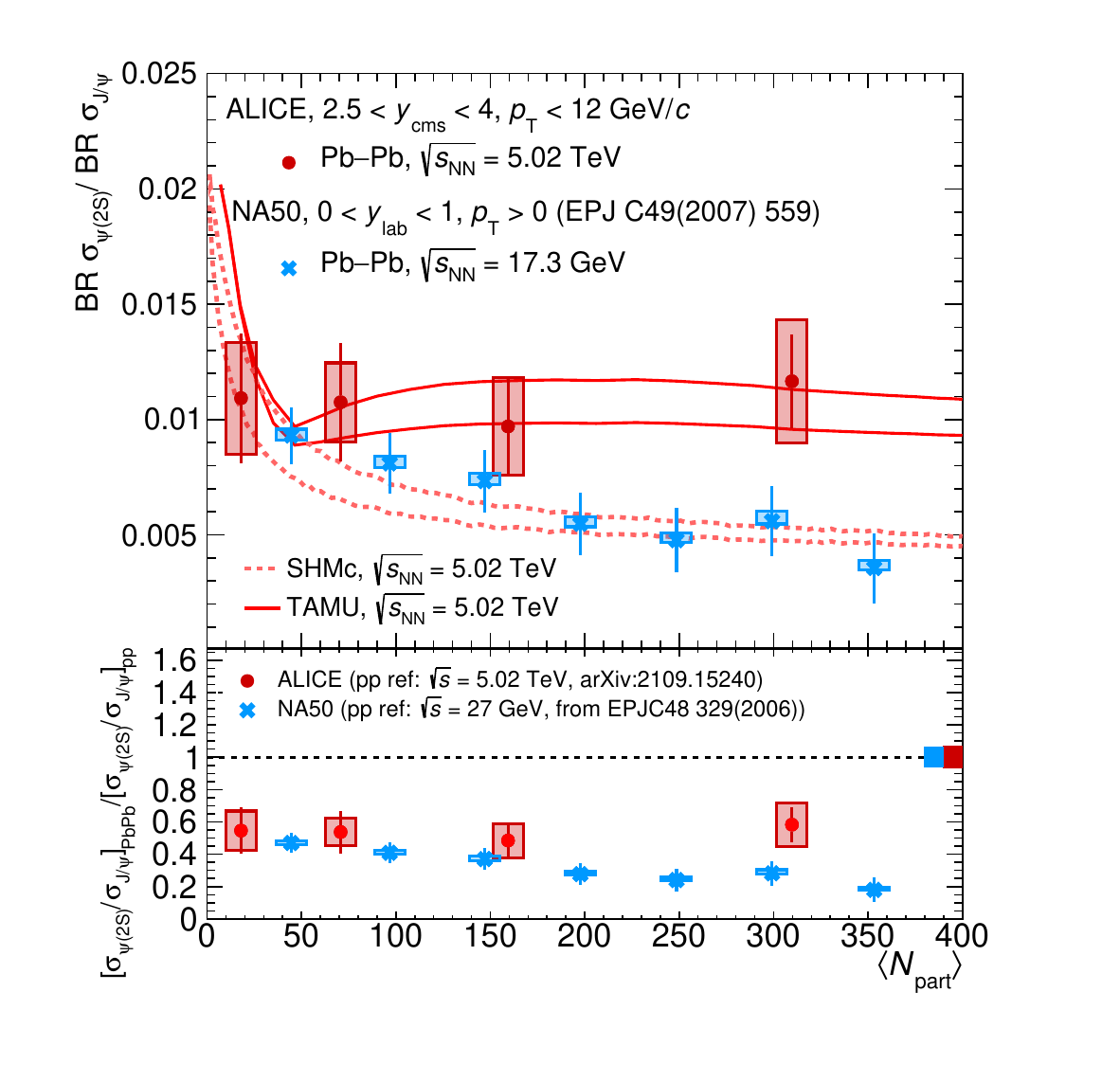}
\end{center}
\caption{Ratio of the $\psi({\rm 2S})$ and J/$\psi$ inclusive cross sections, not corrected for the corresponding branching ratios (BR) of the dimuon decay, as a function of $\langle N_{\rm part}\rangle$. 
The vertical error bars and the filled boxes represent statistical and systematic uncertainties, respectively. Data are compared to predictions of the TAMU~\cite{Du:2015wha} and SHMc~\cite{Andronic:2017pug,Andronic:2019wva} models, the corresponding lines showing their uncertainty band, and to results of the SPS NA50 experiment~\cite{NA50:2006yzz}. In the lower panel, the ratios are normalized to the corresponding pp value (double ratio). The filled boxes around the line at unity indicate the global systematic uncertainties.}
\label{fig:2}
\end{figure}

In Fig.~\ref{fig:2}, the ratio of $\psi({\rm 2S})$ and J/$\psi$ cross sections (not corrected for the branching ratios of the dimuon decay) is shown as a function of centrality, expressed as the average number of participant nucleons $\langle N_{\rm part}\rangle$.  In the lower panel, the values of the double ratio can be read, showing a $\psi({\rm 2S})$ suppression, with respect to J/$\psi$, by a factor of about 2 from pp to Pb--Pb. No significant centrality dependence of the results is seen, within the uncertainties. Comparison  with calculations of a transport approach (TAMU)~\cite{Du:2015wha} and of the SHMc model~\cite{Andronic:2017pug,Andronic:2019wva} are also shown. The TAMU model reproduces the cross section ratios over centrality, while the SHMc model tends to underestimate the data in central Pb--Pb collisions. The ALICE results are also compared with the corresponding inclusive (double) ratios obtained by NA50 in $0<y<1$~\cite{NA50:2006yzz}, which reach smaller values for central events. 

\begin{figure}[t!]
\begin{center}
\includegraphics[scale=0.45]{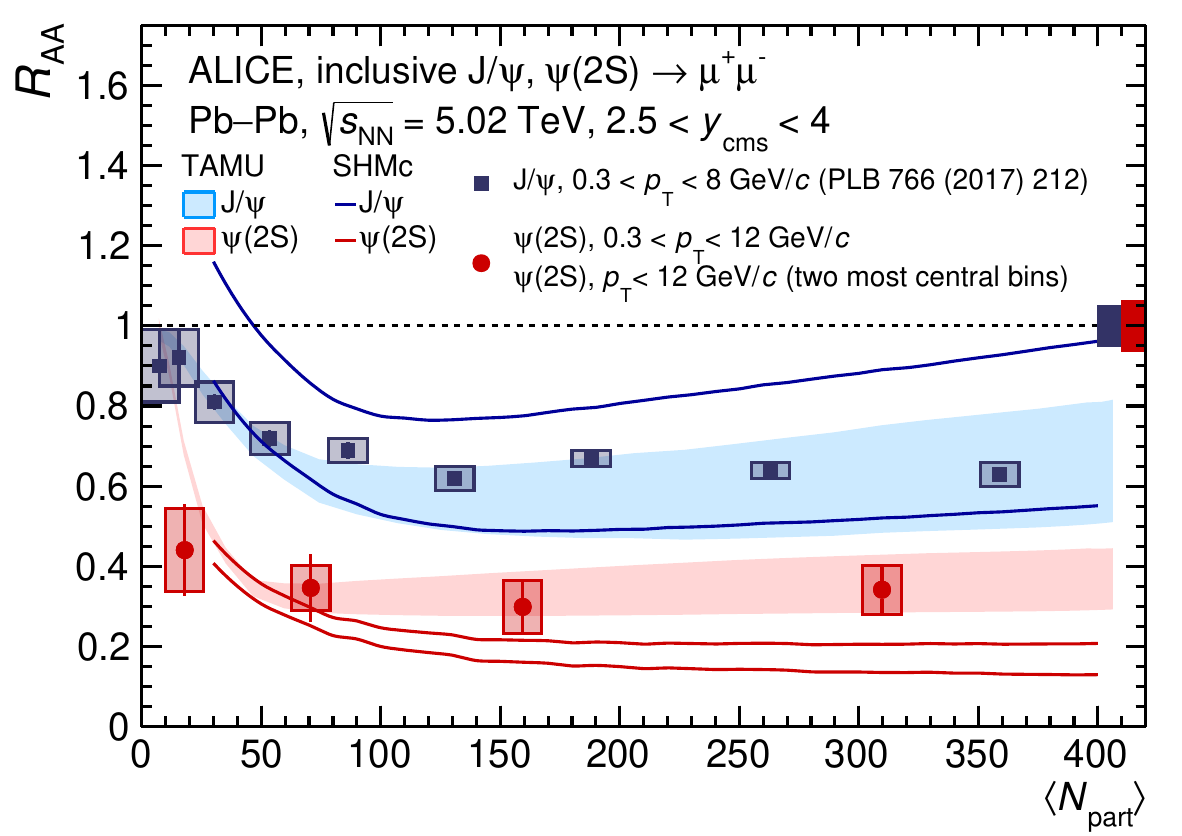}
\end{center}
\caption{The $R_{\rm AA}$ for the $\psi({\rm 2S})$ (this analysis) and J/$\psi$~\cite{Adam:2016rdg} as a function of $\langle N_{\rm part} \rangle$. Comparisons with models are also shown, the lines (band) showing the theory uncertainty for the SHMc (transport) calculations.}
\label{fig:3}
\end{figure}

In Fig.~\ref{fig:3}  the nuclear modification factors for $\psi({\rm 2S})$ (this analysis) and J/$\psi$ (from Ref.~\cite{Adam:2016rdg}) are compared, as a function of $\langle N_{\rm part} \rangle$. With the limited number of centrality  intervals that could be defined, the $R_{\rm AA}$ values for the $\psi({\rm 2S})$ do not show a clear trend and are generally consistent with an $R_{\rm AA}$ value of about 0.4. 
In Fig.~\ref{fig:3}, calculations with the TAMU model are also shown, indicating a good agreement with the measured $R_{\rm AA}$ for both J/$\psi$ and  $\psi({\rm 2S})$. The SHMc model reproduces, within uncertainties, the J/$\psi$ $R_{\rm AA}$ centrality dependence, while it underestimates the $\psi({\rm 2S})$ production in central and semi-central collisions.  

\begin{figure}[!ht]
\begin{center}
\includegraphics[scale=0.45]{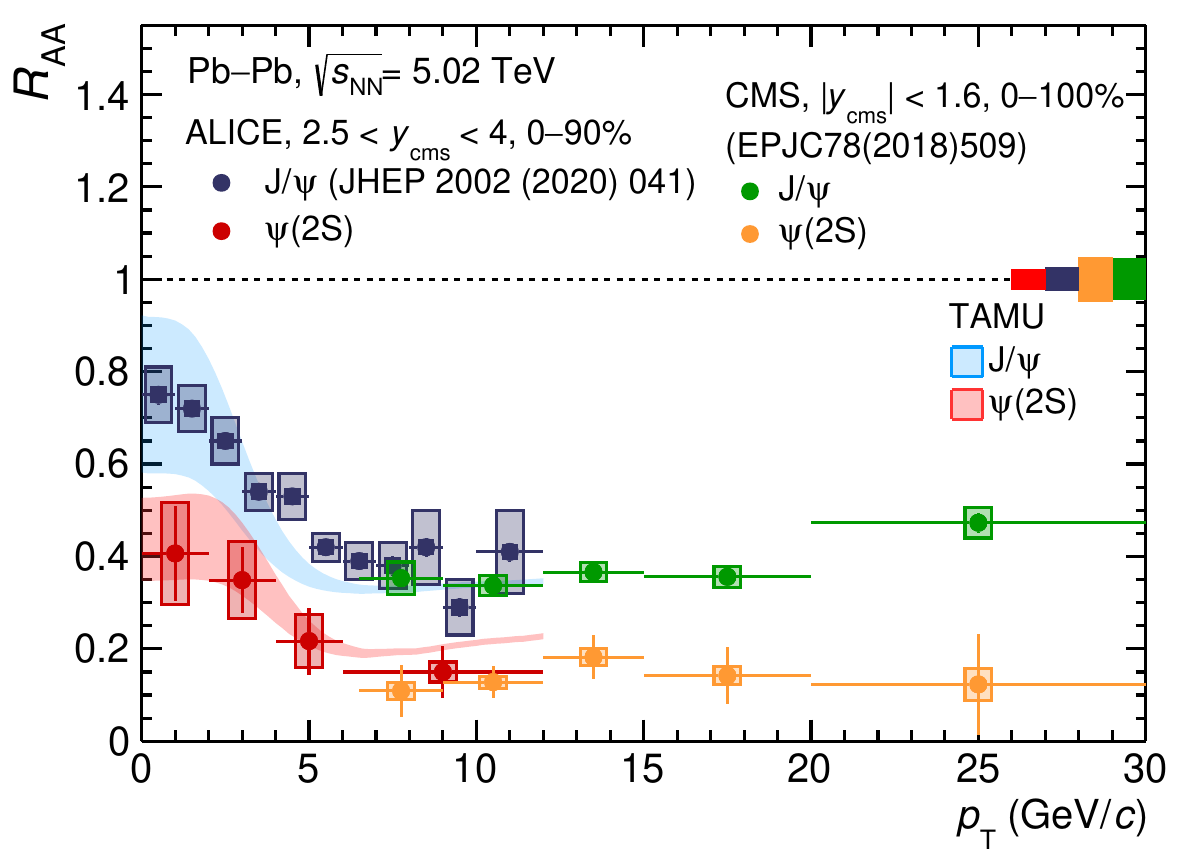}
\end{center}
\caption{The $R_{\rm AA}$ for $\psi({\rm 2S})$ and J/$\psi$~\cite{Adam:2016rdg} as a function of $p_{\rm T}$. Comparison with theory models and results from the CMS experiment~\cite{Sirunyan:2017isk} are also shown.}
\label{fig:5}
\end{figure}

Figure~\ref{fig:5} shows the $\psi({\rm 2S})$ $R_{\rm AA}$, compared with the corresponding result for the J/$\psi$~\cite{Acharya:2019iur}, as a function of $p_{\rm T}$. The corresponding CMS measurements~\cite{CMS:2017uuv} in the region $|y|<1.6$ and $6.5<p_{\rm T}<30$ GeV/$c$ are also reported. The main feature is an increase of the nuclear modification factor at low $p_{\rm T}$, similar to what was observed for the J/$\psi$ and understood as a direct consequence of the recombination process of charm and anticharm quarks. The strong suppression of the $\psi({\rm 2S})$ ($R_{\rm AA}\sim 0.15$ at $p_{\rm T}=10$ GeV/$c$) persists up to $p_{\rm T}=30$ GeV/$c$ as shown by the CMS data, that agree within uncertainties with those of ALICE in the common $p_{\rm T}$ range, in spite of the different rapidity coverage. A comparison with predictions from the TAMU model~\cite{Du:2015wha} is shown, indicating that also the $p_{\rm T}$ dependence  of the $R_{\rm AA}$ is well reproduced for both J/$\psi$ and $\psi({\rm 2S})$, as was the case for the centrality dependence. For completeness, we include in Appendix~\ref{app:doublerat} a plot with the (double) ratio of the $\psi({\rm 2S})$ cross sections as a function of $p_{\rm T}$.

The contribution of nonprompt quarkonia, originating from the decay of b hadrons, could be estimated knowing the fraction $F_{\rm B}$ of nonprompt charmonia in pp collisions, and their $R_{\rm AA}$ in Pb--Pb collisions. An approximate estimate can be carried out using $F_{\rm B}$ values from LHCb~\cite{LHCb:2012geo,LHCb:2021pyk} and the $R_{\rm AA}$ of nonprompt D-mesons measured by ALICE~\cite{ALICE:2022tji} and used as a proxy for nonprompt quarkonia. It shows that prompt  J/$\psi$ and $\psi({\rm 2S})$ $R_{\rm AA}$ would be smaller by less than 1$\sigma$ with respect to their inclusive values. However, it should be noted that the input quantities for this evaluation do not precisely correspond to the same kinematic or collision energy region of our data.

The picture emerging from these results shows a clear hierarchy of suppression of J/$\psi$ and $\psi({\rm 2S})$ over the whole $p_{\rm T}$ and centrality intervals explored. Apart from the overall larger $\psi({\rm 2S})$ suppression, visible in the double ratios and in the comparison of the $R_{\rm AA}$, no significant difference in the $p_{\rm T}$ and centrality dependence of the suppression effects between the two states can be seen. 
The comparison with SPS results shows that the relative suppression of $\psi({\rm 2S})$ with respect to J/$\psi$ in central Pb--Pb events tends to be stronger at low collision energy compared to LHC. However, part of this effect could be due to the different size of the nonprompt component, almost negligible at low energy. 
When comparing the results with the predictions of statistical and transport models, a significantly better agreement is obtained with the latter, in particular for central events, as visible in Figs.~\ref{fig:2} and~\ref{fig:3}. We underline that the data-model comparisons are consistently done for inclusive production. The results shown in this Letter favor the scenario where bound states are dissociated or recombined in the QGP phase, according to the modification of their spectral properties expected from lattice QCD.

In summary, we have provided first accurate results on $\psi({\rm 2S})$ production in Pb--Pb collisions at LHC energy down to zero $p_{\rm T}$, in the rapidity region $2.5<y<4$. Measurements of the cross section ratios between $\psi({\rm 2S})$ and J/$\psi$, of the double ratio between Pb--Pb and pp collisions, and of the nuclear modification factors were shown. A relative suppression by a factor of $\sim 2$ of the $\psi({\rm 2S})$  with respect to the J/$\psi$ is observed, with no significant 
centrality dependence within the  uncertainties. The $R_{\rm AA}$ values for the $\psi({\rm 2S})$ show a hint for a decrease as a function of $p_{\rm T}$, reminiscent of the same effect observed for the J/$\psi$ and connected with charm quark recombination processes. As a function of centrality, values around $R_{\rm AA}\sim 0.4$ are observed. 
The theory comparisons show a good agreement with the predictions of the transport model, that include recombination of charm quarks in the QGP phase, while the SHMc model tends to underestimate the data in central Pb--Pb collisions.

\newenvironment{acknowledgement}{\relax}{\relax}
\begin{acknowledgement}
\section*{Acknowledgements}

The ALICE Collaboration would like to thank all its engineers and technicians for their invaluable contributions to the construction of the experiment and the CERN accelerator teams for the outstanding performance of the LHC complex.
The ALICE Collaboration gratefully acknowledges the resources and support provided by all Grid centres and the Worldwide LHC Computing Grid (WLCG) collaboration.
The ALICE Collaboration acknowledges the following funding agencies for their support in building and running the ALICE detector:
A. I. Alikhanyan National Science Laboratory (Yerevan Physics Institute) Foundation (ANSL), State Committee of Science and World Federation of Scientists (WFS), Armenia;
Austrian Academy of Sciences, Austrian Science Fund (FWF): [M 2467-N36] and Nationalstiftung f\"{u}r Forschung, Technologie und Entwicklung, Austria;
Ministry of Communications and High Technologies, National Nuclear Research Center, Azerbaijan;
Conselho Nacional de Desenvolvimento Cient\'{\i}fico e Tecnol\'{o}gico (CNPq), Financiadora de Estudos e Projetos (Finep), Funda\c{c}\~{a}o de Amparo \`{a} Pesquisa do Estado de S\~{a}o Paulo (FAPESP) and Universidade Federal do Rio Grande do Sul (UFRGS), Brazil;
Bulgarian Ministry of Education and Science, within the National Roadmap for Research Infrastructures 2020-2027 (object CERN), Bulgaria;
Ministry of Education of China (MOEC) , Ministry of Science \& Technology of China (MSTC) and National Natural Science Foundation of China (NSFC), China;
Ministry of Science and Education and Croatian Science Foundation, Croatia;
Centro de Aplicaciones Tecnol\'{o}gicas y Desarrollo Nuclear (CEADEN), Cubaenerg\'{\i}a, Cuba;
Ministry of Education, Youth and Sports of the Czech Republic, Czech Republic;
The Danish Council for Independent Research | Natural Sciences, the VILLUM FONDEN and Danish National Research Foundation (DNRF), Denmark;
Helsinki Institute of Physics (HIP), Finland;
Commissariat \`{a} l'Energie Atomique (CEA) and Institut National de Physique Nucl\'{e}aire et de Physique des Particules (IN2P3) and Centre National de la Recherche Scientifique (CNRS), France;
Bundesministerium f\"{u}r Bildung und Forschung (BMBF) and GSI Helmholtzzentrum f\"{u}r Schwerionenforschung GmbH, Germany;
General Secretariat for Research and Technology, Ministry of Education, Research and Religions, Greece;
National Research, Development and Innovation Office, Hungary;
Department of Atomic Energy Government of India (DAE), Department of Science and Technology, Government of India (DST), University Grants Commission, Government of India (UGC) and Council of Scientific and Industrial Research (CSIR), India;
National Research and Innovation Agency - BRIN, Indonesia;
Istituto Nazionale di Fisica Nucleare (INFN), Italy;
Japanese Ministry of Education, Culture, Sports, Science and Technology (MEXT) and Japan Society for the Promotion of Science (JSPS) KAKENHI, Japan;
Consejo Nacional de Ciencia (CONACYT) y Tecnolog\'{i}a, through Fondo de Cooperaci\'{o}n Internacional en Ciencia y Tecnolog\'{i}a (FONCICYT) and Direcci\'{o}n General de Asuntos del Personal Academico (DGAPA), Mexico;
Nederlandse Organisatie voor Wetenschappelijk Onderzoek (NWO), Netherlands;
The Research Council of Norway, Norway;
Commission on Science and Technology for Sustainable Development in the South (COMSATS), Pakistan;
Pontificia Universidad Cat\'{o}lica del Per\'{u}, Peru;
Ministry of Education and Science, National Science Centre and WUT ID-UB, Poland;
Korea Institute of Science and Technology Information and National Research Foundation of Korea (NRF), Republic of Korea;
Ministry of Education and Scientific Research, Institute of Atomic Physics, Ministry of Research and Innovation and Institute of Atomic Physics and University Politehnica of Bucharest, Romania;
Ministry of Education, Science, Research and Sport of the Slovak Republic, Slovakia;
National Research Foundation of South Africa, South Africa;
Swedish Research Council (VR) and Knut \& Alice Wallenberg Foundation (KAW), Sweden;
European Organization for Nuclear Research, Switzerland;
Suranaree University of Technology (SUT), National Science and Technology Development Agency (NSTDA), Thailand Science Research and Innovation (TSRI) and National Science, Research and Innovation Fund (NSRF), Thailand;
Turkish Energy, Nuclear and Mineral Research Agency (TENMAK), Turkey;
National Academy of  Sciences of Ukraine, Ukraine;
Science and Technology Facilities Council (STFC), United Kingdom;
National Science Foundation of the United States of America (NSF) and United States Department of Energy, Office of Nuclear Physics (DOE NP), United States of America.
In addition, individual groups or members have received support from:
Marie Sk\l{}odowska Curie, European Research Council, Strong 2020 - Horizon 2020 (grant nos. 950692, 824093, 896850), European Union;
Academy of Finland (Center of Excellence in Quark Matter) (grant nos. 346327, 346328), Finland;
Programa de Apoyos para la Superaci\'{o}n del Personal Acad\'{e}mico, UNAM, Mexico.
\end{acknowledgement}

\bibliographystyle{utphys}  
\bibliography{bibliography}


\newpage
\appendix
\section{Invariant-mass plots and fits}
\label{app:fits}

\begin{figure}[h]
    \centering
    \includegraphics[scale=0.31]{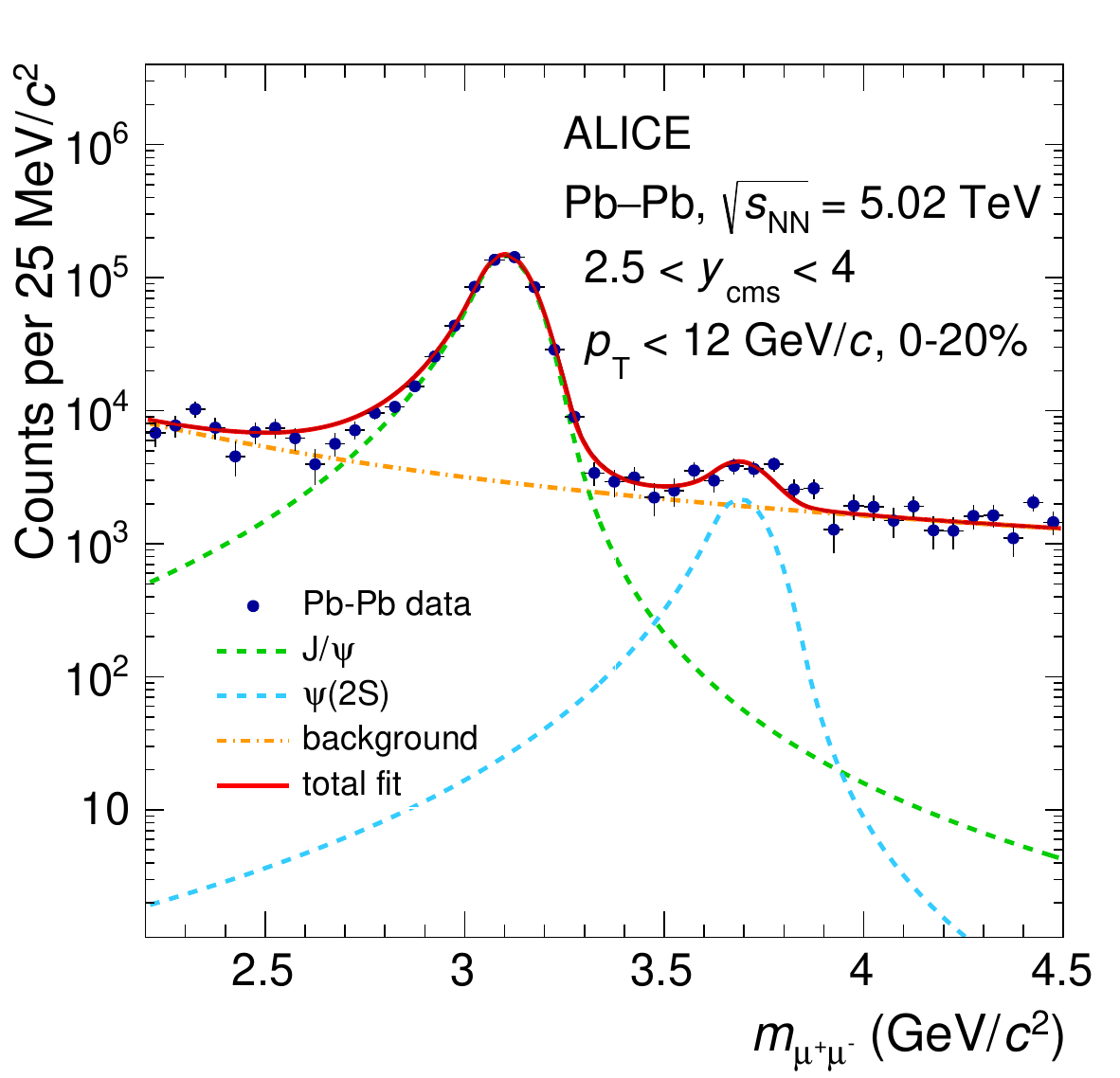}
    \includegraphics[scale=0.31]{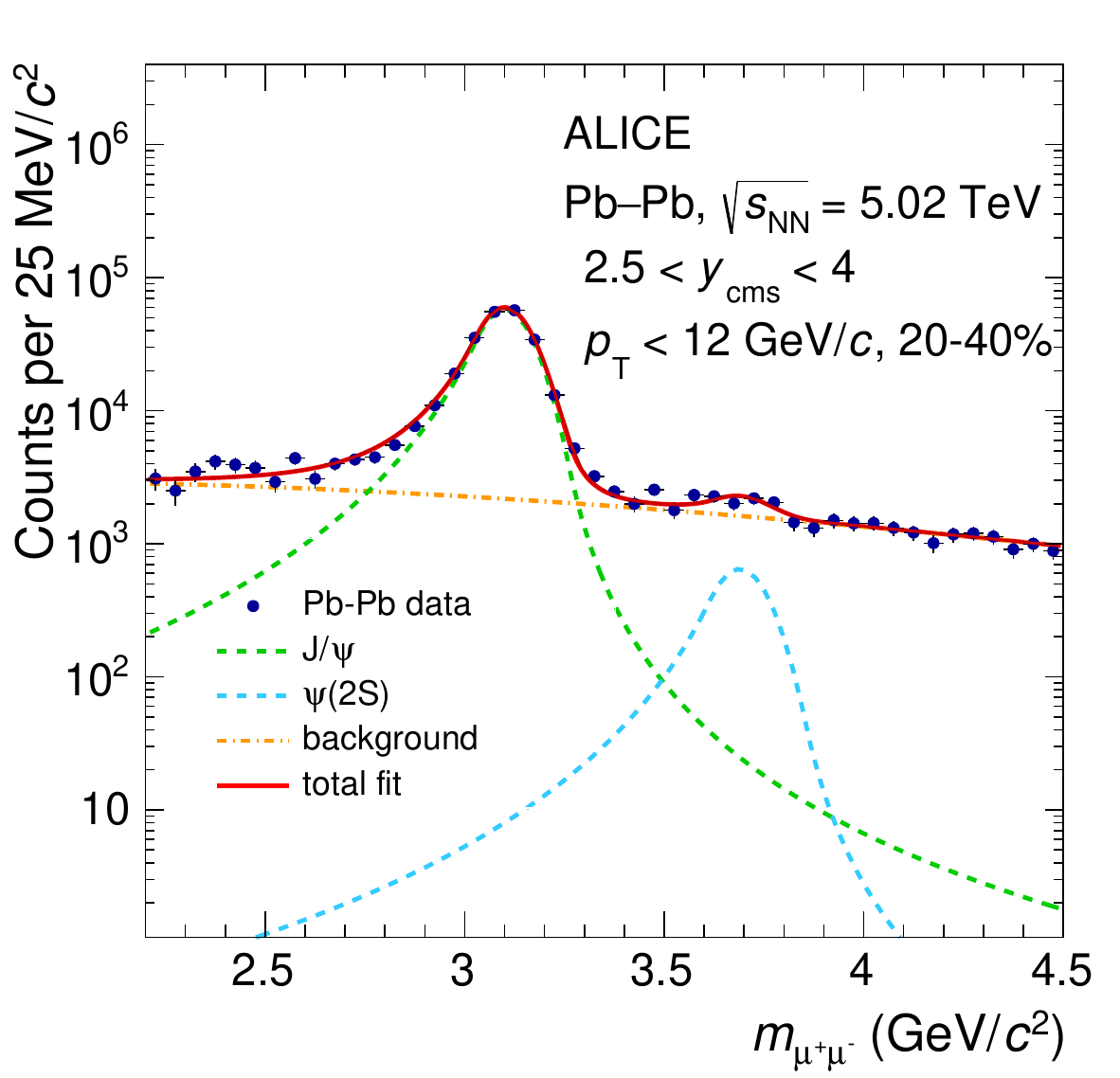}
    \includegraphics[scale=0.31]{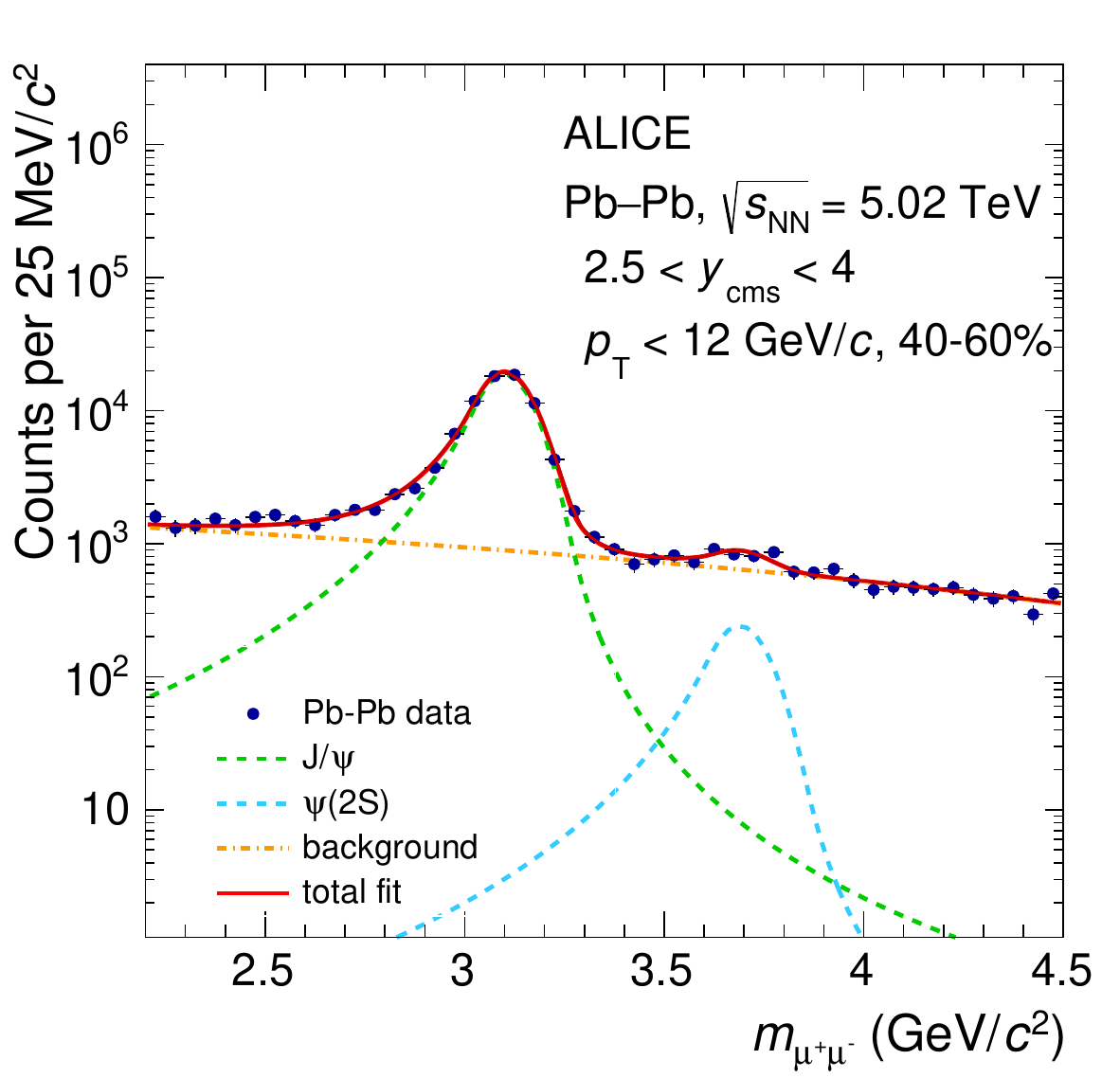}
    \includegraphics[scale=0.31]{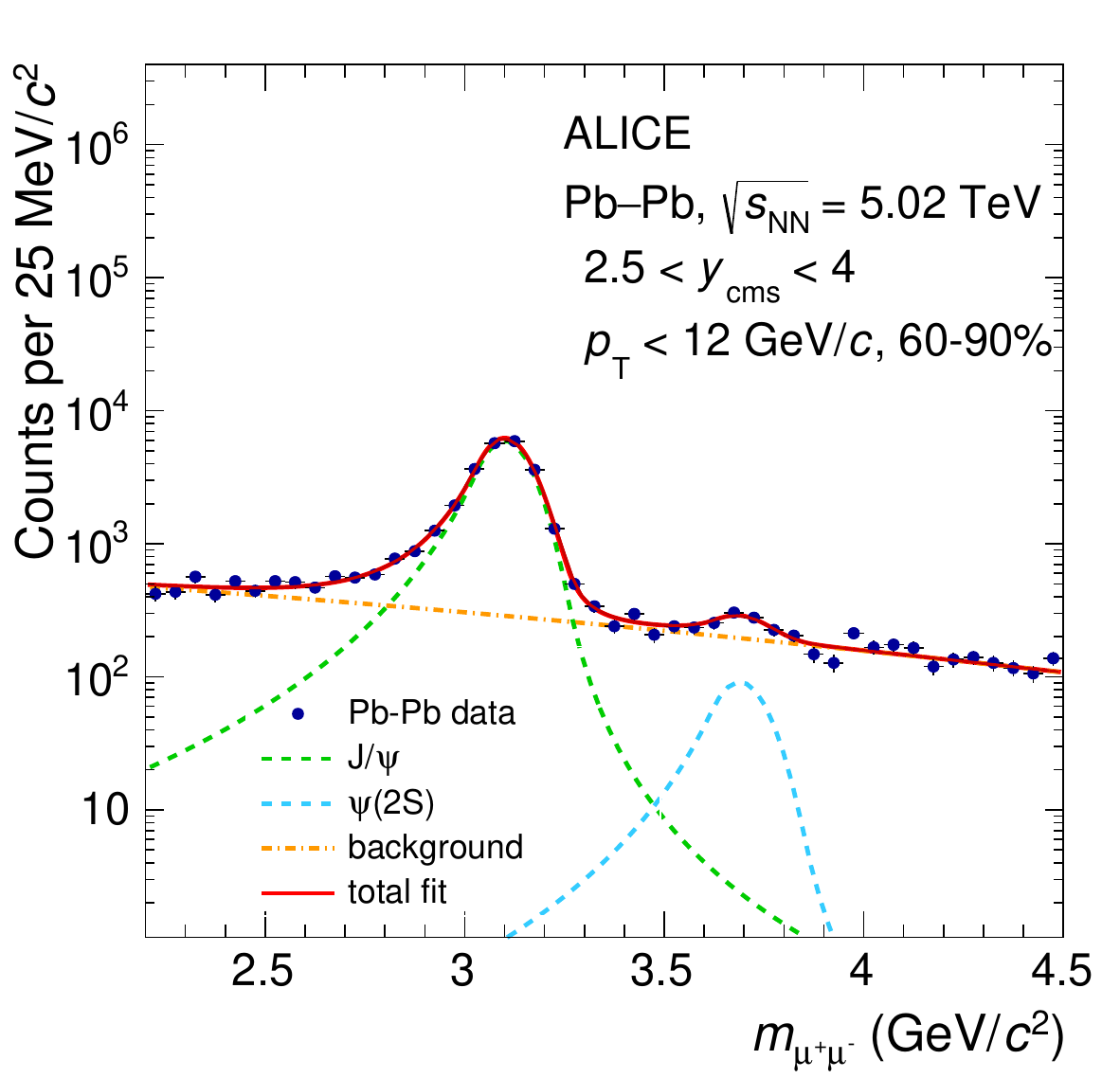}
    \caption{The opposite-sign invariant mass spectrum of mass pairs, for the four centrality intervals used in the data analysis, after mixed-event background subtraction. The kinematic domain is $2.5<y<4$, $p_{\rm T}<12$ GeV/$c$ ($0.3<p_{\rm T}<12$ GeV/$c$ for the two most peripheral intervals). The result of one among the 60 $\chi^2$ minimization fit is reported, corresponding to a CB function for the signal, a variable-width Gaussian for the background, $2<m_{\mu\mu}<5$ GeV/$c^2$ fitting range, a $2<m_{\mu\mu}<8$ GeV/$c^2$ normalization range of the mixed-event background, and CB tails obtained from Monte Carlo. The contribution of the J/$\psi$ and $\psi(2S)$ resonances and of the background continuum as obtained from the fit are also reported.}
    \label{fig:invmassvscent}
\end{figure}

\begin{figure}[h]
   \centering
    \includegraphics[scale=0.31]{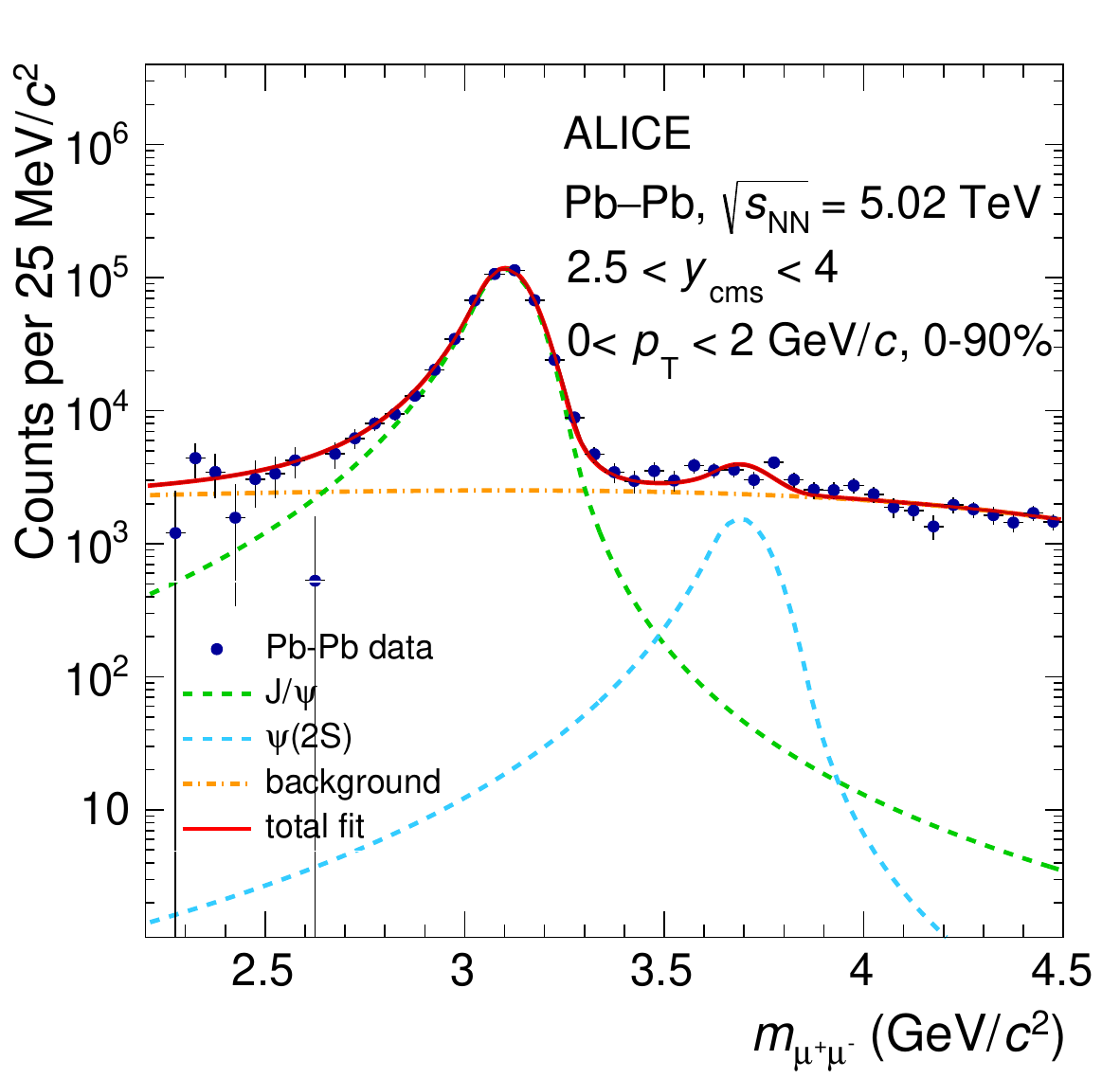}
    \includegraphics[scale=0.31]{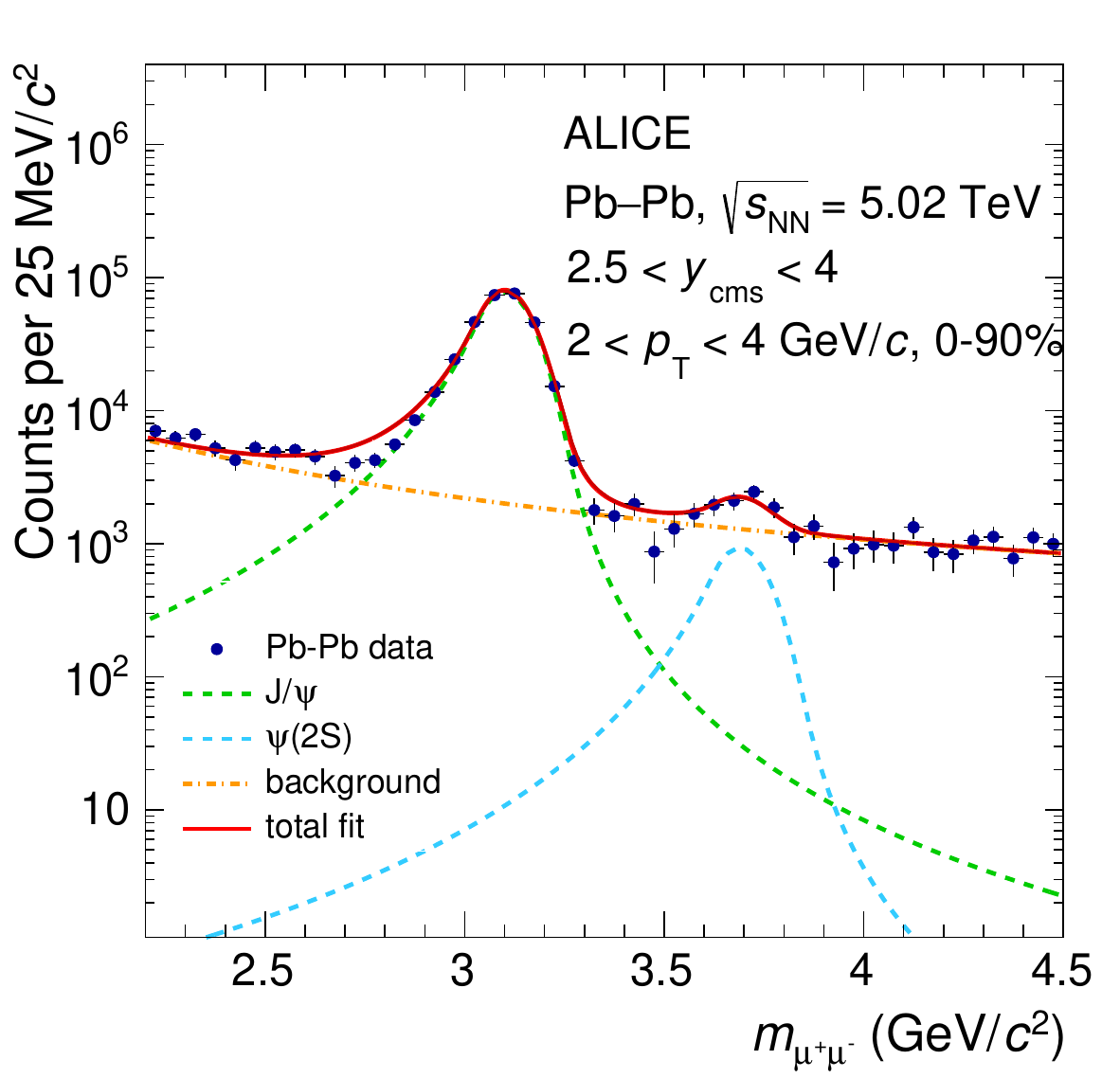}
    \includegraphics[scale=0.31]{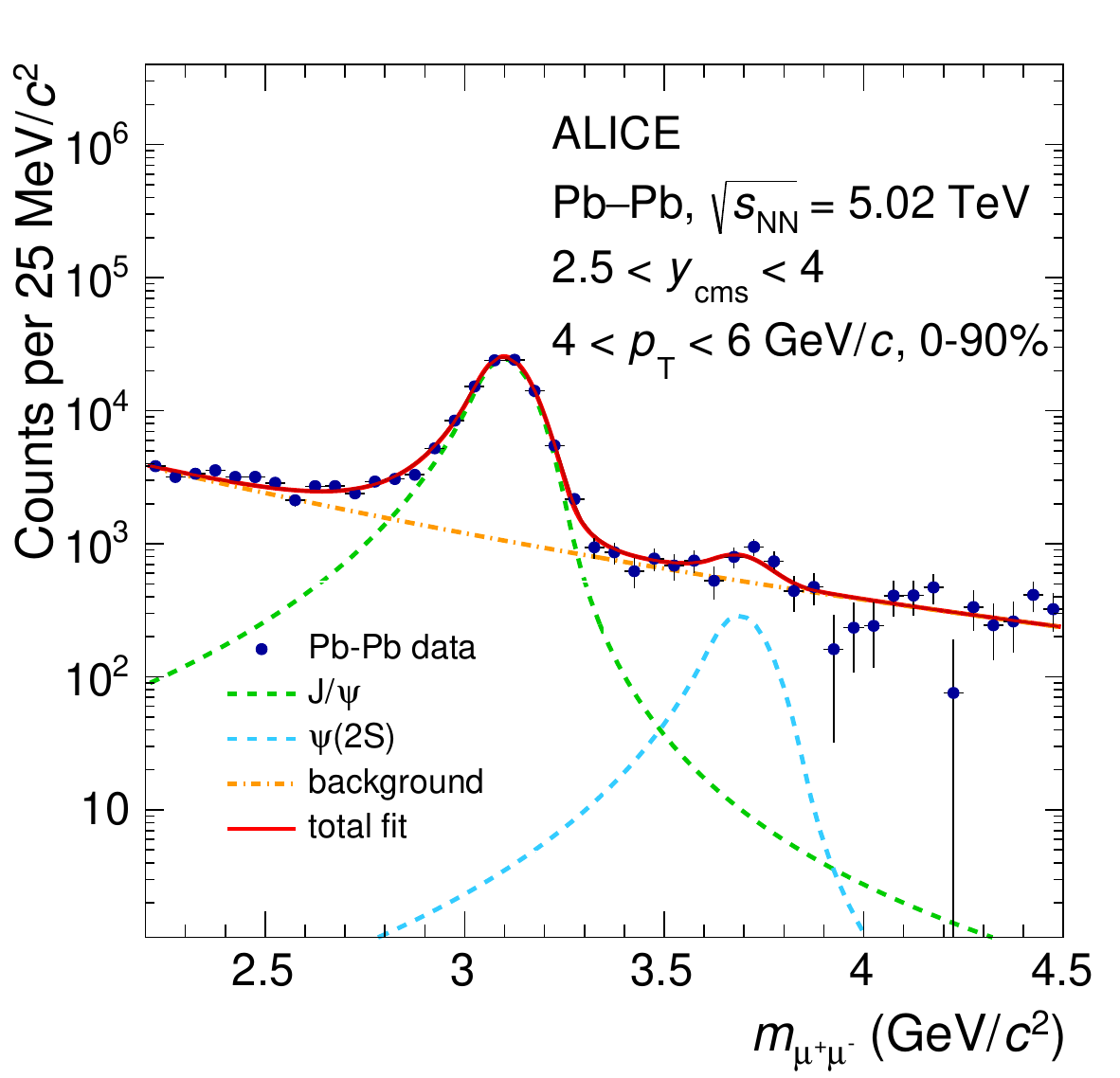}
    \includegraphics[scale=0.31]{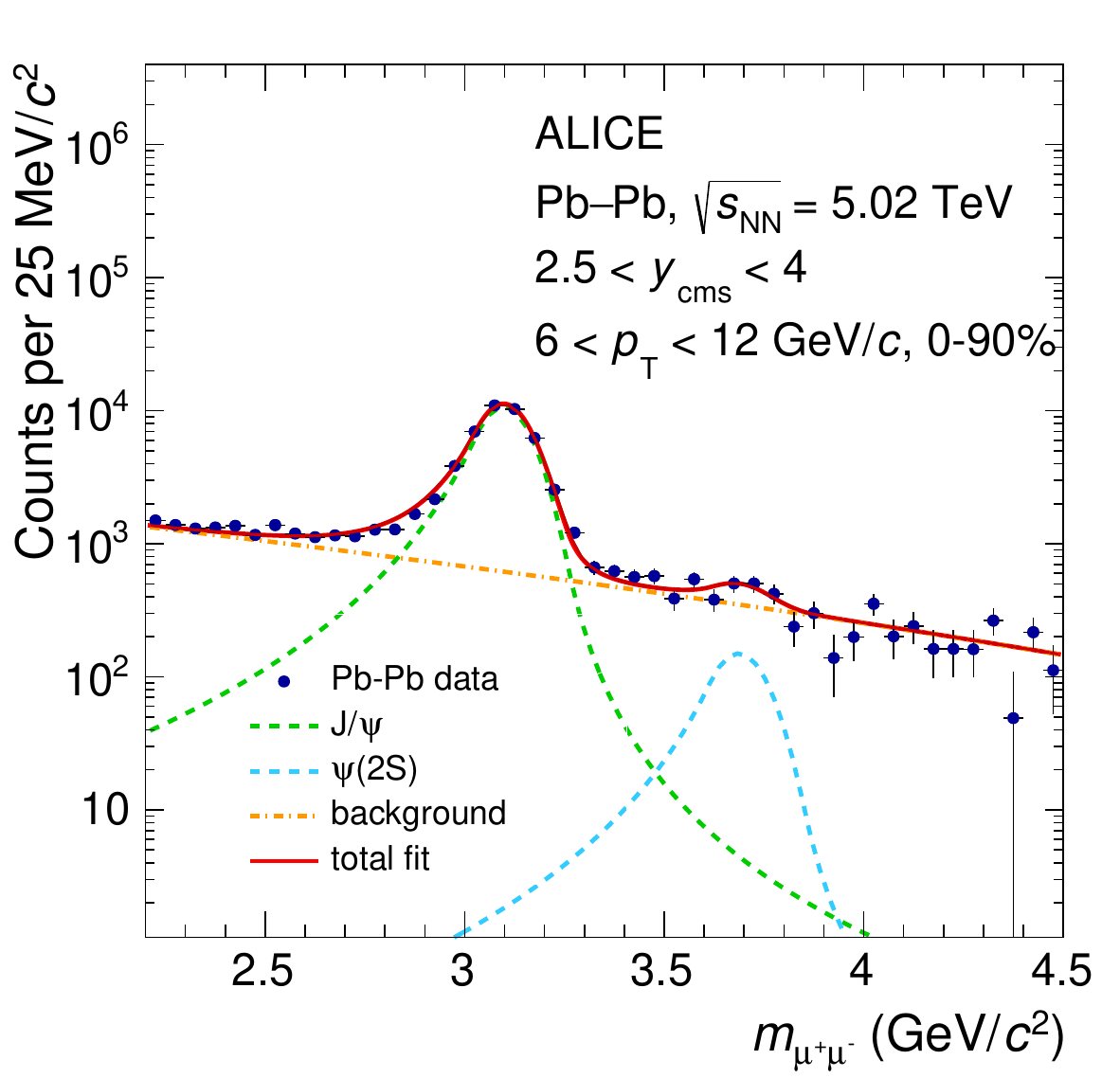}
    \caption{The opposite-sign invariant mass spectrum of mass pairs, for the four transverse momentum intervals used in the data analysis, after mixed-event background subtraction. The kinematic domain is $2.5<y<4$ and the centrality interval is 0--90\%. The result of one among the 60 $\chi^2$ minimization fit is reported, corresponding to a CB function for the signal, a variable-width Gaussian for the background, a $2<m_{\mu\mu}<5$ GeV/$c^2$ fitting range, $2<m_{\mu\mu}<8$ GeV/$c^2$ normalization range of the mixed-event background, and CB tails obtained from Monte Carlo. The contribution of the J/$\psi$ and $\psi(2S)$ resonances and of the background continuum as obtained from the fit are also reported.}
    \label{fig:invmassvspt}
\end{figure}

The dimuon invariant mass distributions, after mixed-event background subtraction, are reported in Fig.~\ref{fig:invmassvscent} and Fig.~\ref{fig:invmassvspt}, together with the result of fits used to extract the number of J/$\psi$ and $\psi(2S)$. Detailed information is hereafter given on the fit inputs, further details on the procedure are reported in the Letter.  

For the fits as a function of centrality, 60 variations of the fit inputs were considered. They correspond to: (i) two different choices for the resonance shapes, either Crystal Ball (CB) function or a pseudo-Gaussian with a mass-dependent width~\cite{ALICE-Quarkonia-signal-extraction}; (ii)  five possible choices for the background shape, a variable-width Gaussian, an exponential or the sum of two exponential functions, a 1$^{\rm st}$ or 2$^{\rm nd}$ degree polynomial; (iii) two choices of the fitting range, either $2.2<m_{\mu\mu}<4.5$ GeV/$c^2$ or $2<m_{\mu\mu}<5$ GeV/$c^2$; (iv) two choices for the normalization range of the mixed-event background, either $2<m_{\mu\mu}<8$ GeV/$c^2$ or $2.5<m_{\mu\mu}<7$ GeV/$c^2$; (v) two choices for the tails of the CB function, either from Monte Carlo or from a high-statistics data sample collected in pp collisions at 13 TeV. 
The minimum  significance  of the $\psi({\rm 2S})$ signal as a function of centrality is 7.6 (60--90\%) and increases by a factor $\sim$ 8 for central events. 
Numerical values on $\psi({\rm 2S})$-related quantities are given in Table~\ref{tab:sm}.

\newpage
For the fits as a function of transverse momentum, 36 variations of the fit inputs were considered, They  correspond to: (i) two different choices for the resonance shapes, either Crystal Ball (CB) function or a pseudo-Gaussian with a mass-dependent width~\cite{ALICE-Quarkonia-signal-extraction}; (ii) three possible choices for the background shape, a variable-width Gaussian, an exponential, a 2$^{\rm nd}$ degree polynomial; (iii) two choices of the fitting range, either $2.2<m_{\mu\mu}<4.5$ GeV/$c^2$ or $2<m_{\mu\mu}<5$ GeV/$c^2$; (iv) two choices for the normalization range of the mixed-event background, either $2<m_{\mu\mu}<8$ GeV/$c^2$ or $2.5<m_{\mu\mu}<7$ GeV/$c^2$; (v) two choices for the tails of the CB function, either from Monte Carlo or from a high-statistics data sample collected in pp collisions at 13 TeV. The minimum  significance  of the $\psi({\rm 2S})$ signal as a function of $p_{\rm T}$ is 11.3 at high $p_{\rm T}$, with an increase of a factor $\sim 3$ at low $p_{\rm T}$.

\begin{table}[!ht]
	\begin{center}
		\caption{With reference to Figs.~\ref{fig:invmassvscent} and ~\ref{fig:invmassvspt}, number of $\psi({\rm 2S})$ signal counts and background counts in a $3\sigma$ interval around the $\psi({\rm 2S})$ mass. The relative statistical uncertainty on the final number of $\psi({\rm 2S})$ is also listed.}\label{tab:sm}
		\begin{tabular}{c|c|c|c}
			\hline
			Centrality & $N_{\rm sig}$ & $N_{\rm bck}$ & Stat. unc. (\%)\\
			\hline
            0--20\% & 9675 & 14211 & 17.7 \\
            \hline
            20--40\% & 2474 & 13917 & 22.9 \\
            \hline
            40--60\% & 928 & 5530 & 23.2 \\
            \hline
            60--90\% & 342 & 1645 & 21.6 \\
            \hline
            $p_{\rm T}$ (GeV/$c$) & & &\\
			\hline
            0--2 & 4342 & 22223 & 22.4 \\
            \hline
            2--4 & 3494 & 10791 & 17.9 \\
            \hline
            4--6 & 1094 & 4541 & 30.4 \\
            \hline
            6--12 & 567 & 2996 & 33.6 \\

		\end{tabular}
	\end{center}
\end{table}

\newpage

\section{$p_{\rm T}$ dependence of the (double) ratios of the $\psi({\rm 2S})$ and J/$\psi$ cross sections}
\label{app:doublerat}

\begin{figure}[!ht]
\begin{center}
\includegraphics[scale=0.50]{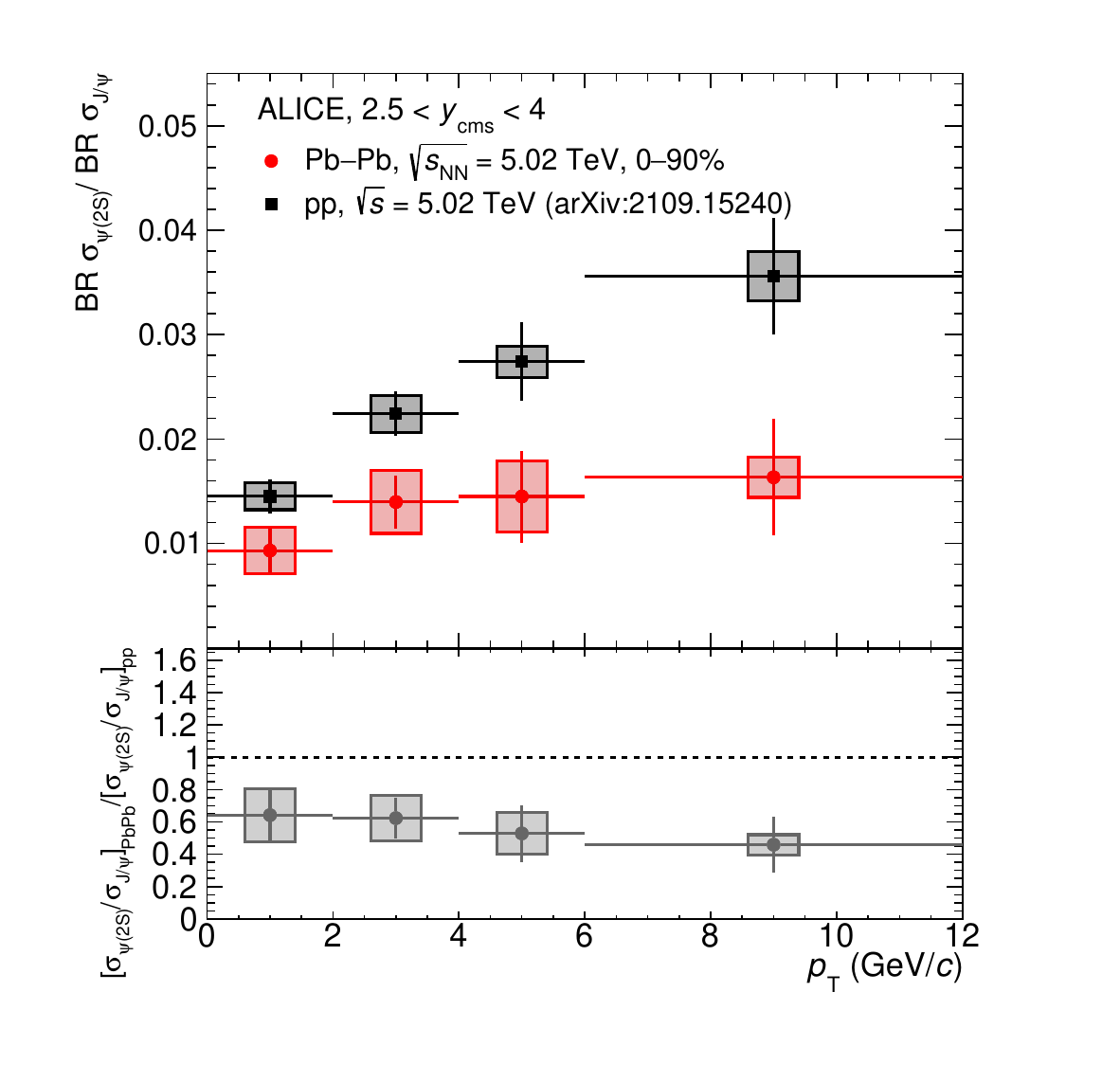}
\end{center}
\caption{The ratio of the $\psi({\rm 2S})$ and J/$\psi$ cross sections as a function of $p_{\rm T}$, not corrected for the corresponding branching ratios BR of the dimuon decay, compared to measurements in pp collisions~\cite{ALICE:2021qlw}. In the lower panel the double ratios of the Pb--Pb and pp values are shown.}
\label{fig:4}
\end{figure}

In Fig.~\ref{fig:4} the $p_{\rm T}$-dependence of the ratio of the $\psi({\rm 2S})$ and J/$\psi$ production cross sections for Pb--Pb collisions, in 0--90\% centrality,  is presented and compared to the corresponding pp results. The pp ratios increase as a function of $p_{\rm T}$, while for Pb--Pb no significant rise is present. The double ratio values are also shown, indicating a significant relative suppression of the $\psi({\rm 2S})$, slightly increasing with $p_{\rm T}$ up to a value of $\sim 0.5$.

\clearpage

\newpage

\section{The ALICE Collaboration}
\label{app:collab}
\begin{flushleft} 
\small

S.~Acharya\,\orcidlink{0000-0002-9213-5329}\,$^{\rm 125}$, 
D.~Adamov\'{a}\,\orcidlink{0000-0002-0504-7428}\,$^{\rm 86}$, 
A.~Adler$^{\rm 69}$, 
G.~Aglieri Rinella\,\orcidlink{0000-0002-9611-3696}\,$^{\rm 32}$, 
M.~Agnello\,\orcidlink{0000-0002-0760-5075}\,$^{\rm 29}$, 
N.~Agrawal\,\orcidlink{0000-0003-0348-9836}\,$^{\rm 50}$, 
Z.~Ahammed\,\orcidlink{0000-0001-5241-7412}\,$^{\rm 132}$, 
S.~Ahmad\,\orcidlink{0000-0003-0497-5705}\,$^{\rm 15}$, 
S.U.~Ahn\,\orcidlink{0000-0001-8847-489X}\,$^{\rm 70}$, 
I.~Ahuja\,\orcidlink{0000-0002-4417-1392}\,$^{\rm 37}$
% ...

\section*{Affiliation Notes}

$^{\rm I}$ Deceased\\
$^{\rm II}$ Also at: Max-Planck-Institut f\"{u}r Physik, Munich, Germany\\
$^{\rm III}$ Also at: Italian National Agency for New Technologies, Energy and Sustainable Economic Development (ENEA), Bologna, Italy\\
$^{\rm IV}$ Also at: Dipartimento DET del Politecnico di Torino, Turin, Italy\\
$^{\rm V}$ Also at: Department of Applied Physics, Aligarh Muslim University, Aligarh, India\\
$^{\rm VI}$ Also at: Institute of Theoretical Physics, University of Wroclaw, Poland\\
$^{\rm VII}$ Also at: An institution covered by a cooperation agreement with CERN\\

\section*{Collaboration Institutes}

$^{1}$ A.I. Alikhanyan National Science Laboratory (Yerevan Physics Institute) Foundation, Yerevan, Armenia\\
$^{2}$ AGH University of Science and Technology, Cracow, Poland\\
$^{3}$ Bogolyubov Institute for Theoretical Physics, National Academy of Sciences of Ukraine, Kiev, Ukraine\\
$^{4}$ Bose Institute, Department of Physics  and Centre for Astroparticle Physics and Space Science (CAPSS), Kolkata, India\\
$^{5}$ California Polytechnic State University, San Luis Obispo, California, United States\\
$^{6}$ Central China Normal University, Wuhan, China\\
$^{7}$ Centro de Aplicaciones Tecnol\'{o}gicas y Desarrollo Nuclear (CEADEN), Havana, Cuba\\
$^{8}$ Centro de Investigaci\'{o}n y de Estudios Avanzados (CINVESTAV), Mexico City and M\'{e}rida, Mexico\\
$^{9}$ Chicago State University, Chicago, Illinois, United States\\
$^{10}$ China Institute of Atomic Energy, Beijing, China\\
$^{11}$ Chungbuk National University, Cheongju, Republic of Korea\\
$^{12}$ Comenius University Bratislava, Faculty of Mathematics, Physics and Informatics, Bratislava, Slovak Republic\\
$^{13}$ COMSATS University Islamabad, Islamabad, Pakistan\\
$^{14}$ Creighton University, Omaha, Nebraska, United States\\
$^{15}$ Department of Physics, Aligarh Muslim University, Aligarh, India\\
$^{16}$ Department of Physics, Pusan National University, Pusan, Republic of Korea\\
$^{17}$ Department of Physics, Sejong University, Seoul, Republic of Korea\\
$^{18}$ Department of Physics, University of California, Berkeley, California, United States\\
$^{19}$ Department of Physics, University of Oslo, Oslo, Norway\\
$^{20}$ Department of Physics and Technology, University of Bergen, Bergen, Norway\\
$^{21}$ Dipartimento di Fisica, Universit\`{a} di Pavia, Pavia, Italy\\
$^{22}$ Dipartimento di Fisica dell'Universit\`{a} and Sezione INFN, Cagliari, Italy\\
$^{23}$ Dipartimento di Fisica dell'Universit\`{a} and Sezione INFN, Trieste, Italy\\
$^{24}$ Dipartimento di Fisica dell'Universit\`{a} and Sezione INFN, Turin, Italy\\
$^{25}$ Dipartimento di Fisica e Astronomia dell'Universit\`{a} and Sezione INFN, Bologna, Italy\\
$^{26}$ Dipartimento di Fisica e Astronomia dell'Universit\`{a} and Sezione INFN, Catania, Italy\\
$^{27}$ Dipartimento di Fisica e Astronomia dell'Universit\`{a} and Sezione INFN, Padova, Italy\\
$^{28}$ Dipartimento di Fisica `E.R.~Caianiello' dell'Universit\`{a} and Gruppo Collegato INFN, Salerno, Italy\\
$^{29}$ Dipartimento DISAT del Politecnico and Sezione INFN, Turin, Italy\\
$^{30}$ Dipartimento di Scienze MIFT, Universit\`{a} di Messina, Messina, Italy\\
$^{31}$ Dipartimento Interateneo di Fisica `M.~Merlin' and Sezione INFN, Bari, Italy\\
$^{32}$ European Organization for Nuclear Research (CERN), Geneva, Switzerland\\
$^{33}$ Faculty of Electrical Engineering, Mechanical Engineering and Naval Architecture, University of Split, Split, Croatia\\
$^{34}$ Faculty of Engineering and Science, Western Norway University of Applied Sciences, Bergen, Norway\\
$^{35}$ Faculty of Nuclear Sciences and Physical Engineering, Czech Technical University in Prague, Prague, Czech Republic\\
$^{36}$ Faculty of Physics, Sofia University, Sofia, Bulgaria\\
$^{37}$ Faculty of Science, P.J.~\v{S}af\'{a}rik University, Ko\v{s}ice, Slovak Republic\\
$^{38}$ Frankfurt Institute for Advanced Studies, Johann Wolfgang Goethe-Universit\"{a}t Frankfurt, Frankfurt, Germany\\
$^{39}$ Fudan University, Shanghai, China\\
$^{40}$ Gangneung-Wonju National University, Gangneung, Republic of Korea\\
$^{41}$ Gauhati University, Department of Physics, Guwahati, India\\
$^{42}$ Helmholtz-Institut f\"{u}r Strahlen- und Kernphysik, Rheinische Friedrich-Wilhelms-Universit\"{a}t Bonn, Bonn, Germany\\
$^{43}$ Helsinki Institute of Physics (HIP), Helsinki, Finland\\
$^{44}$ High Energy Physics Group,  Universidad Aut\'{o}noma de Puebla, Puebla, Mexico\\
$^{45}$ Horia Hulubei National Institute of Physics and Nuclear Engineering, Bucharest, Romania\\
$^{46}$ Indian Institute of Technology Bombay (IIT), Mumbai, India\\
$^{47}$ Indian Institute of Technology Indore, Indore, India\\
$^{48}$ INFN, Laboratori Nazionali di Frascati, Frascati, Italy\\
$^{49}$ INFN, Sezione di Bari, Bari, Italy\\
$^{50}$ INFN, Sezione di Bologna, Bologna, Italy\\
$^{51}$ INFN, Sezione di Cagliari, Cagliari, Italy\\
$^{52}$ INFN, Sezione di Catania, Catania, Italy\\
$^{53}$ INFN, Sezione di Padova, Padova, Italy\\
$^{54}$ INFN, Sezione di Pavia, Pavia, Italy\\
$^{55}$ INFN, Sezione di Torino, Turin, Italy\\
$^{56}$ INFN, Sezione di Trieste, Trieste, Italy\\
$^{57}$ Inha University, Incheon, Republic of Korea\\
$^{58}$ Institute for Gravitational and Subatomic Physics (GRASP), Utrecht University/Nikhef, Utrecht, Netherlands\\
$^{59}$ Institute of Experimental Physics, Slovak Academy of Sciences, Ko\v{s}ice, Slovak Republic\\
$^{60}$ Institute of Physics, Homi Bhabha National Institute, Bhubaneswar, India\\
$^{61}$ Institute of Physics of the Czech Academy of Sciences, Prague, Czech Republic\\
$^{62}$ Institute of Space Science (ISS), Bucharest, Romania\\
$^{63}$ Institut f\"{u}r Kernphysik, Johann Wolfgang Goethe-Universit\"{a}t Frankfurt, Frankfurt, Germany\\
$^{64}$ Instituto de Ciencias Nucleares, Universidad Nacional Aut\'{o}noma de M\'{e}xico, Mexico City, Mexico\\
$^{65}$ Instituto de F\'{i}sica, Universidade Federal do Rio Grande do Sul (UFRGS), Porto Alegre, Brazil\\
$^{66}$ Instituto de F\'{\i}sica, Universidad Nacional Aut\'{o}noma de M\'{e}xico, Mexico City, Mexico\\
$^{67}$ iThemba LABS, National Research Foundation, Somerset West, South Africa\\
$^{68}$ Jeonbuk National University, Jeonju, Republic of Korea\\
$^{69}$ Johann-Wolfgang-Goethe Universit\"{a}t Frankfurt Institut f\"{u}r Informatik, Fachbereich Informatik und Mathematik, Frankfurt, Germany\\
$^{70}$ Korea Institute of Science and Technology Information, Daejeon, Republic of Korea\\
$^{71}$ KTO Karatay University, Konya, Turkey\\
$^{72}$ Laboratoire de Physique des 2 Infinis, Ir\`{e}ne Joliot-Curie, Orsay, France\\
$^{73}$ Laboratoire de Physique Subatomique et de Cosmologie, Universit\'{e} Grenoble-Alpes, CNRS-IN2P3, Grenoble, France\\
$^{74}$ Lawrence Berkeley National Laboratory, Berkeley, California, United States\\
$^{75}$ Lund University Department of Physics, Division of Particle Physics, Lund, Sweden\\
$^{76}$ Nagasaki Institute of Applied Science, Nagasaki, Japan\\
$^{77}$ Nara Women{'}s University (NWU), Nara, Japan\\
$^{78}$ National and Kapodistrian University of Athens, School of Science, Department of Physics , Athens, Greece\\
$^{79}$ National Centre for Nuclear Research, Warsaw, Poland\\
$^{80}$ National Institute of Science Education and Research, Homi Bhabha National Institute, Jatni, India\\
$^{81}$ National Nuclear Research Center, Baku, Azerbaijan\\
$^{82}$ National Research and Innovation Agency - BRIN, Jakarta, Indonesia\\
$^{83}$ Niels Bohr Institute, University of Copenhagen, Copenhagen, Denmark\\
$^{84}$ Nikhef, National institute for subatomic physics, Amsterdam, Netherlands\\
$^{85}$ Nuclear Physics Group, STFC Daresbury Laboratory, Daresbury, United Kingdom\\
$^{86}$ Nuclear Physics Institute of the Czech Academy of Sciences, Husinec-\v{R}e\v{z}, Czech Republic\\
$^{87}$ Oak Ridge National Laboratory, Oak Ridge, Tennessee, United States\\
$^{88}$ Ohio State University, Columbus, Ohio, United States\\
$^{89}$ Physics department, Faculty of science, University of Zagreb, Zagreb, Croatia\\
$^{90}$ Physics Department, Panjab University, Chandigarh, India\\
$^{91}$ Physics Department, University of Jammu, Jammu, India\\
$^{92}$ Physics Program and International Institute for Sustainability with Knotted Chiral Meta Matter (SKCM2), Hiroshima University, Hiroshima, Japan\\
$^{93}$ Physikalisches Institut, Eberhard-Karls-Universit\"{a}t T\"{u}bingen, T\"{u}bingen, Germany\\
$^{94}$ Physikalisches Institut, Ruprecht-Karls-Universit\"{a}t Heidelberg, Heidelberg, Germany\\
$^{95}$ Physik Department, Technische Universit\"{a}t M\"{u}nchen, Munich, Germany\\
$^{96}$ Politecnico di Bari and Sezione INFN, Bari, Italy\\
$^{97}$ Research Division and ExtreMe Matter Institute EMMI, GSI Helmholtzzentrum f\"ur Schwerionenforschung GmbH, Darmstadt, Germany\\
$^{98}$ Saga University, Saga, Japan\\
$^{99}$ Saha Institute of Nuclear Physics, Homi Bhabha National Institute, Kolkata, India\\
$^{100}$ School of Physics and Astronomy, University of Birmingham, Birmingham, United Kingdom\\
$^{101}$ Secci\'{o}n F\'{\i}sica, Departamento de Ciencias, Pontificia Universidad Cat\'{o}lica del Per\'{u}, Lima, Peru\\
$^{102}$ Stefan Meyer Institut f\"{u}r Subatomare Physik (SMI), Vienna, Austria\\
$^{103}$ SUBATECH, IMT Atlantique, Nantes Universit\'{e}, CNRS-IN2P3, Nantes, France\\
$^{104}$ Sungkyunkwan University, Suwon City, Republic of Korea\\
$^{105}$ Suranaree University of Technology, Nakhon Ratchasima, Thailand\\
$^{106}$ Technical University of Ko\v{s}ice, Ko\v{s}ice, Slovak Republic\\
$^{107}$ The Henryk Niewodniczanski Institute of Nuclear Physics, Polish Academy of Sciences, Cracow, Poland\\
$^{108}$ The University of Texas at Austin, Austin, Texas, United States\\
$^{109}$ Universidad Aut\'{o}noma de Sinaloa, Culiac\'{a}n, Mexico\\
$^{110}$ Universidade de S\~{a}o Paulo (USP), S\~{a}o Paulo, Brazil\\
$^{111}$ Universidade Estadual de Campinas (UNICAMP), Campinas, Brazil\\
$^{112}$ Universidade Federal do ABC, Santo Andre, Brazil\\
$^{113}$ University of Cape Town, Cape Town, South Africa\\
$^{114}$ University of Houston, Houston, Texas, United States\\
$^{115}$ University of Jyv\"{a}skyl\"{a}, Jyv\"{a}skyl\"{a}, Finland\\
$^{116}$ University of Kansas, Lawrence, Kansas, United States\\
$^{117}$ University of Liverpool, Liverpool, United Kingdom\\
$^{118}$ University of Science and Technology of China, Hefei, China\\
$^{119}$ University of South-Eastern Norway, Kongsberg, Norway\\
$^{120}$ University of Tennessee, Knoxville, Tennessee, United States\\
$^{121}$ University of the Witwatersrand, Johannesburg, South Africa\\
$^{122}$ University of Tokyo, Tokyo, Japan\\
$^{123}$ University of Tsukuba, Tsukuba, Japan\\
$^{124}$ University Politehnica of Bucharest, Bucharest, Romania\\
$^{125}$ Universit\'{e} Clermont Auvergne, CNRS/IN2P3, LPC, Clermont-Ferrand, France\\
$^{126}$ Universit\'{e} de Lyon, CNRS/IN2P3, Institut de Physique des 2 Infinis de Lyon, Lyon, France\\
$^{127}$ Universit\'{e} de Strasbourg, CNRS, IPHC UMR 7178, F-67000 Strasbourg, France, Strasbourg, France\\
$^{128}$ Universit\'{e} Paris-Saclay Centre d'Etudes de Saclay (CEA), IRFU, D\'{e}partment de Physique Nucl\'{e}aire (DPhN), Saclay, France\\
$^{129}$ Universit\`{a} degli Studi di Foggia, Foggia, Italy\\
$^{130}$ Universit\`{a} del Piemonte Orientale, Vercelli, Italy\\
$^{131}$ Universit\`{a} di Brescia, Brescia, Italy\\
$^{132}$ Variable Energy Cyclotron Centre, Homi Bhabha National Institute, Kolkata, India\\
$^{133}$ Warsaw University of Technology, Warsaw, Poland\\
$^{134}$ Wayne State University, Detroit, Michigan, United States\\
$^{135}$ Westf\"{a}lische Wilhelms-Universit\"{a}t M\"{u}nster, Institut f\"{u}r Kernphysik, M\"{u}nster, Germany\\
$^{136}$ Wigner Research Centre for Physics, Budapest, Hungary\\
$^{137}$ Yale University, New Haven, Connecticut, United States\\
$^{138}$ Yonsei University, Seoul, Republic of Korea\\
$^{139}$  Zentrum  f\"{u}r Technologie und Transfer (ZTT), Worms, Germany\\
$^{140}$ Affiliated with an institute covered by a cooperation agreement with CERN\\
$^{141}$ Affiliated with an international laboratory covered by a cooperation agreement with CERN.\\

\end{flushleft} 

\end{document}